# Fabrication of thin planar radiopure foils with $^{82}$Se for the SuperNEMO Demonstrator

X. Aguerre,[a,b] A. Barabash,[c] A. Basharina-Freshville,[d] M. Bongrand,[e] Ch. Bourgeois,[e] D. Breton,[e] R. Breier,[f] J. Busto,[g] C. Cerna,[a] J. Cesar,[h] M. Ceschia,[d] E. Chauveau,[a] S. De Capua,[i] D. Duchesneau,[j] J.J. Evans,[i] D.V. Filosofov,[c] M. Granjon,[a] M. Hoballah,[e] R. Hodák,[k] J. Horkley,[l] A. Jeremie,[j] S. Jullian,[e] J. Kaizer,[f] A.A. Klimenko,[c] O. Kochetov,[c] F. Koňařík,[k,m] S. Konovalov,[c] T. Křižák,[k,m] A. Lahaie,[a] K. Lang,[h] Y. Lemière,[n] T. Le Noblet,[j] P. Li,[b] P. Loaiza,[e] J. Maalmi,[e] M. Macko,[k]   F. Mamedov,[k] C. Marquet,[a] F. Mauger,[n] A. Mendl,[k,o] B. Morgan,[p] I. Nemchenok,[c] V. Palušová,[k] C. Patrick,[b] F. Perrot,[a] M. Petro,[f,k] F. Piquemal,[a] P. Povinec,[f] S. Pratt,[b] M. Proga,[h] W.S. Quinn,[d] A.V. Rakhimov,[c] Y. Ramachers,[p] A. Remoto,[j] N.I. Rukhadze,[c] R. Saakyan,[d] R. Salazar,[h] J. Sedgbeer,[q] Y. Shitov,[k] L. Simard,[e] F. Šimkovic,[f,k] A.A. Smolnikov,[c] S. Söldner-Rembold,[j,q] I. Štekl,[k] J. Suhonen,[r] H. Tedjditi,[g] J. Thomas,[d] V. Timkin,[c] V. Tretyak,[c] V.I. Tretyak,[s,t] G. Turnbull,[b] Y. Vereshchaka,[e] G. Warot,[u] D. Waters[d] and V. Yumatov[c]

[a]Université de Bordeaux, CNRS/IN2P3, LP2i, Bordeaux, UMR 5797, F-33170, Gradignan, France
[b]University of Edinburgh, SUPA, School of Physics and Astronomy, Edinburgh, EH9 3FD, United Kingdom
[c]Participant in the NEMO-3/SuperNEMO collaboration
[d]University College London, London, WC1E 6BT, United Kingdom
[e]Université Paris-Saclay, CNRS, IJCLab, F-91405, Orsay, France
[f]Faculty of Mathematics, Physics and Informatics, Comenius University, SK-842 48, Bratislava, Slovakia
[g]Aix-Marseille Université, CNRS, CPPM, F-13288 Marseille, France
[h]University of Texas at Austin, Department of Physics, Austin, TX 78712, USA
[j]University of Manchester, Manchester, M13 9PL, United Kingdom
[i]LAPP, Université Savoie Mont Blanc, CNRS/IN2P3, Annecy, France
[k]Institute of Experimental and Applied Physics, Czech Technical University in Prague, CZ-11000 Prague, Czech Republic
[l]Idaho National Laboratory, Idaho Falls, ID 83415, USA
[m]Faculty of Nuclear Sciences and Physical Engineering, Czech Technical University in Prague, 115 19 Prague, Czech Republic
[n]Université de Caen Normandie, ENSICAEN, CNRS/IN2P3, LPC Caen, UMR6534, F-14000 Caen, France
[o]Faculty of Mathematics and Physics, Charles University, CZ-12116 Prague, Czech Republic
[p]University of Warwick, Coventry, CV4 7AL, United Kingdom
[q]Imperial College London, London, SW7 2BZ, United Kingdom
[r]Department of Physics, University of Jyväskylä, Jyväskylä, Finland
[s]Institute for Nuclear Research of NASU, Kyiv, 03028, Ukraine
[t]INFN, Laboratori Nazionali del Gran Sasso, 67100, Assergi (AQ), Italy
[u]Univ. Grenoble Alpes, CNRS, Grenoble INP, LPSC-IN2P3, 38000 Grenoble, France.

*Abstract:*

*The SuperNEMO Demonstrator, designed to search for double beta decay using enriched $^{82}$Se, has been assembled in the Modane Underground Laboratory under the French Alps. Thin foils with radio-*

*purified and enriched $^{82}$Se are installed centrally in the detector. A novel foil fabrication method has been developed, improving the radiopurity achieved in the previous generation experiment. It consists of wrapping standalone selenium pads in raw Mylar, combined with selenium purified by a new reverse-chromatography method. This paper describes the features of these foils, their fabrication process, the characterization results, and the integration of the foils into the SuperNEMO Demonstrator.*

# 1. Introduction

The SuperNEMO experiment aims to detect neutrinoless double beta decay (0νββ), a hypothetical process that would violate lepton number conservation by two units. According to the Schechter-Valle theorem [1], the observation of such a decay would prove that neutrinos are massive Majorana particles, i.e., they are their own antiparticles, regardless of the underlying mechanism. Majorana neutrinos could explain the mass of the neutrino via the seesaw mechanism [2] and help to explain the matter-antimatter asymmetry via the leptogenesis process [3]. The most stringent limits on the half-life of 0νββ decay have been set at up to $10^{26}$ years [4, 5].

In this context, the SuperNEMO Demonstrator has been installed in the Modane Underground Laboratory (LSM) under the Alps [6]. This Demonstrator serves as a proof-of-concept demonstration for a potential larger experiment to study 0νββ mechanisms in the event of a discovery with the capability to integrate multiple similar modules [7]. In addition, the large size of the Demonstrator allows it to be used for conducting physics investigations.

A plane of thin foils with enriched $^{82}$Se, a potential source of a neutrinoless double beta decay, is installed centrally in the detector. Radiopurity of this plane is central to the physics reach of the experiment. It is crucial to minimize the presence of unwanted radioactive contaminants that can mimic 0νββ signature. We discuss detailed steps how to produce the foils and our effort to characterize them.

# 2. SuperNEMO Demonstrator, general description

The SuperNEMO Demonstrator employs a pioneering tracking-calorimetry technique, inherited from the NEMO-3 experiment [8]. This method combines a thin, independent double beta (ββ) decay source with a tracking volume that records the trajectories of charged particles. Surrounding this tracking volume is a segmented calorimeter that measures the energy and timing of individual particles. One key advantage of this approach is that the source is physically separated from the detector. This allows for flexibility in the choice of isotope and enables easy replacement with a different isotope if needed. Additionally, the technique offers particle identification and full topological reconstruction of the two-electron signature characteristic of ββ decay. This enables effective background suppression and potential discrimination between different double beta decay mechanisms by analyzing the angular distribution and individual energies of the emitted electrons [7].

These capabilities also allow for detailed topological studies of two-neutrino double beta (2νββ) events, which provide valuable input for nuclear calculations and searches for physics beyond the Standard Model. Special attention has been given to minimizing background sources to maximize the detector's sensitivity to the 0νββ decay process.

Figure 1 shows the SuperNEMO Demonstrator along with its overall dimensions. The experiment uses 6.11 kg of $^{82}$Se as the ββ source.

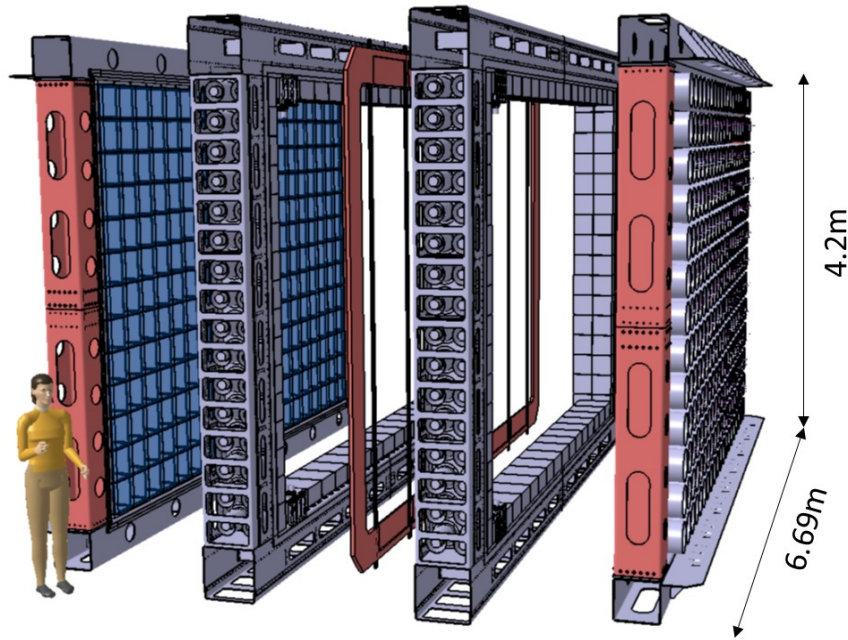

*Figure 1: Exploded view of the SuperNEMO Demonstrator with dimensions. In the centre is the source foil frame, sandwiched between the two tracker volumes and the two calorimeter walls.*

## 2.1 ββ Source

The ββ source located at the center of the Demonstrator consists of 34 selenium foils and 2 copper foils. Copper, known for its ultra-radiopurity, is used in SuperNEMO as a control to help quantify background levels arising from radioactivity external to the source foils. The total mass of selenium in the Demonstrator is 6.25 kg, including 6.11 kg of $^{82}$Se, compared to 0.35 kg of copper.

To produce radiopure $^{82}$Se, six different purification processes were studied [9]. Each foil measures 2700 × 135 × 0.30 mm, yielding a total surface area of 13 m². The foils have a surface density of 40–60 mg/cm² and an enrichment level of 96%–99.9% in $^{82}$Se.

To evaluate different foil preparation techniques, the 34 selenium foils were produced using two distinct methods. Nineteen foils were fabricated using the method developed for NEMO-3, in which a selenium powder mixture is sandwiched between two long strips of 12 μm-thick microperforated Mylar, spanning the full height of the detector. The remaining 15 foils were produced using a newly develo-ped technique, involving shorter, standalone selenium pads that were later assembled and wrapped in 12 μm-thick raw Mylar.

Both preparation methods are described in detail in Section 3.

## 2.2 Tracker

The tracking chamber comprises 2034 drift cells operating in Geiger mode organized into 2 chambers, each containing 113 rows of 9 cells. It represents an assembly of 14970 anode and cathode wires installed with the help of a wiring robot. The tracker is filled with a mixture of 95% helium, 4% ethanol ($C_2H_5OH$) as a quencher, and 1% argon; it isregulated by a dedicated gas-mixing system [10].

## 2.3 Calorimeter

The calorimeter is composed of 712 Optical Modules (OMs) made out of polystyrene-based scintillator (except for 8 OMs using polyvinyltoluene (PVT) scintillators), and Hamamatsu 5" or 8" photomultiplier tubes (PMTs). Each OM is wrapped in Teflon and Mylar to optimize light reflection, and surrounded by an individual pure-iron magnetic shield. The 520 PMTs forming the two main walls parallel to the selenium foils are directly coupled to the scintillator, for optimal light collection. A mean energy resolution of 8.3% FWHM at 1 MeV was achieved for the main wall OMs equipped with 8" PMTs exposed to an electron beam [11] during construction tests [12,13].

## 2.4 Background reduction

An experiment's sensitivity to the half-life of rare processes such as neutrinoless double beta decay varies inversely with the level of background. Therefore, background suppression is critical to maximizing sensitivity.

Several sources of background are considered and can be categorized as either internal or external:

- **Internal background** arises from natural radioactive contamination within the ββ source foils themselves, primarily from isotopes such as $^{208}$Tl and $^{214}$Bi.

- **External background** originates from neutrons and γ-rays produced by the surrounding laboratory environment or the detector structure, as well as from radon contamination in the laboratory air.

A range of strategies has been implemented to reduce and monitor these backgrounds:

- **Material screening**: Selenium, foil components, and materials used during the fabrication process were screened with high-purity germanium (HPGe) detectors, achieving sensitivity levels in the range of 0.1–1 mBq/kg.
- **Quantifying internal contamination**: The radioactive content of the source foils was measured using a dedicated detector, BiPo-3, which operated between 2013 and 2018 at the Canfranc Underground Laboratory (LSC) in Spain. BiPo-3 achieved sensitivity levels of 140 μBq/kg for $^{214}$Bi and 2 μBq/kg for $^{208}$Tl [14].
- **Independent verification**: Selenium samples were also analyzed using Inductively Coupled Plasma Mass Spectrometry (ICP-MS) for cross-validation.
- **Passive shielding**: Dedicated γ and neutron shieldings have been installed around the detector to minimize external background radiation.
- **Radon mitigation**: Radon emanation and permeability were controlled through careful material selection and extensive testing using dedicated setups, including a radon concentration line capable of detecting extremely low radon levels. An anti-radon tent made from black po-lycarbonate surrounds the Demonstrator and is flushed with radon-free air supplied by a facility equipped with a radon trap. These measures aim to maintain radon activity within the detector below 0.15 mBq/m³.
- **Magnetic field application**: A magnetic coil surrounds the detector, generating a field of approximately 25 Gauss. This field allows for the discrimination between electrons and positrons, particularly those produced via pair production by high-energy γ-rays resulting from neutron capture events within the detector.

## 2.5 Calibration and monitoring

Absolute energy calibration is accomplished by a system of 42 $^{207}$Bi radioactive sources that can be automatically deployed between the source foils in the middle of the Demonstrator. Figure 2 shows a diagram of the selenium and copper foils in place on their frame together with the $^{207}$Bi calibration lines. The $^{207}$Bi sources sources have been thoroughly characterized [15], and their activities measured with sub-percent precision ranging from 125 Bq to 145 Bq [16]. These sources are deployed periodically between running periods. For fast and daily optical module response calibration, a light-injection system is used to monitor any relative drift of the calorimeter's response to within 1% precision. There are 20 pulsed UV LEDs that can inject light into calorimeter modules via optical fibers. Five additional reference optical modules, with permanent $^{241}$Am and $^{207}$Bi sources, are used to control the stability of the LED light [17]. For time-resolution measurements, $^{60}$Co sources can also be installed in the deployment system, alongside $^{207}$Bi sources.

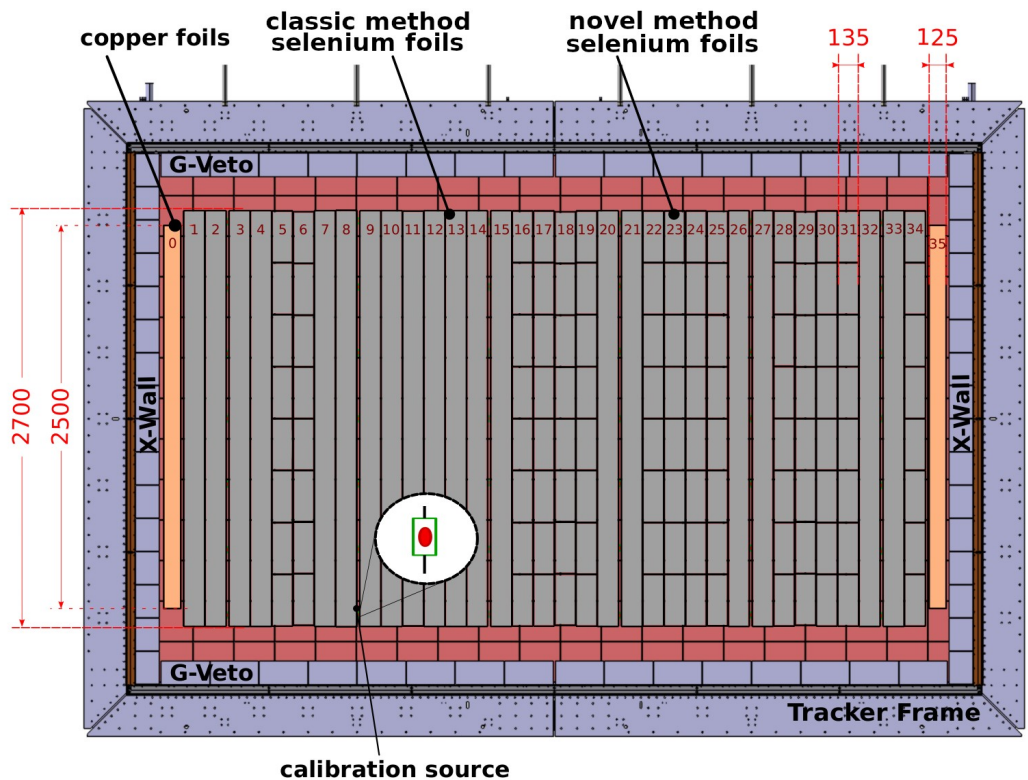

Figure 2: 36 ⸺ source foils attached on the source frame: 34 ββ selenium foils (in grey) and 2 radiopure copper (in orange) foils at the extremities along with the 6 calibration lines (vertical black lines), each equipped with 7 $^{207}$Bi sources (white inset as an example). Dimensions are given in mm.

# 3. Foil fabrication

## 3.1 Choice of the isotope and design of the source foils

A key design feature of SuperNEMO is that its sensitive components (tracker and calorimeter) are physically separate from the ββ source. This allows the use of any solid isotope, unlike other experiments where the source must also serve as the calorimeter.

Isotope selection prioritized the ββ transition energy $Q_{\beta\beta}$, the total energy of the two electrons in 0νββ decay. Since the decay rate is proportional to $Q_{\beta\beta}^5$, higher $Q_{\beta\beta}$ values increase sensitivity for a given run time and help suppress backgrounds from the 2.615 MeV γ-ray of [208]Tl decay and the 3.27 MeV β-decay of [214]Bi in the radon chain. Isotopes with $Q_{\beta\beta}$ above these energies are therefore preferred.

Only isotopes with natural abundance above 2% were considered, as enrichment costs decrease with increasing abundance.

Finally, because the Standard Model decay to two neutrinos, 2νββ, is an irreducible background to the beyond the Standard Model 0νββ process, this background can be reduced by choosing an isotope with a long 2νββ half-life. A table of transition energies for various ββ isotopes is provided in reference [18].

Considering these reasons, [82]Se ($Q_{\beta\beta}$= 2.995 MeV, $T_{1/2}^{2\nu}$=8.60 ± 0.03(stat.) $^{+0.19}_{-0.13}$ (syst.) × 10$^{19}$ yr [19]) was chosen for SuperNEMO.

The source foil thickness was selected to maximize experimental sensitivity by balancing the largest possible isotope mass against β-particle inefficiency and energy loss upon exiting the foil. Thicker foils increase the likelihood of internal scattering, thereby degrading energy resolution. The thickness was therefore optimized so that its contribution to the overall energy resolution matched that of the next most significant factor - the calorimeter resolution.

The dominant internal background arises from surface contamination, mainly from radon-daughter decay, which becomes proportionally more significant for thinner foils due to increased surface area. For SuperNEMO's calorimeter resolution and expected background levels, the optimal surface density is 40–60 mg/cm², corresponding to a maximum foil thickness of about 300 μm [8].

## 3.2 Foil preparation methods

The source foils were prepared using two methods. The method used in NEMO-3 is well proven. A novel method, expected to be more radiopure, was also tested. Both methods will be described below. All materials used in foil production were screened for radiopurity levels.

### – The NEMO-3 method

The first method used for the SuperNEMO foil fabrication was the one used in NEMO-3 [8] for the composite foils. The selenium is mixed with polyvinyl alcohol (PVA) glue, and enclosed in 12 μm-thick microperforated Mylar called *backing film*. This film is produced by first irradiating it with a [84]Kr ion beam of 3 MeV/nucleon to produce holes at a density of 10$^7$ per cm². Chemical etching is then performed using NaOH (5M) at 70°C, after which the foils are washed with 1% acetic acid and dried in hot air.

For foil preparation, pure selenium powder was used. The first step involved sieving the selenium powder to keep grains smaller than 45 µm. Any larger grains were ground up and several additional sieving and grinding processes were undertaken until the grains were small enough to ensure a good bond to the backing film. Next, the powder was mixed with the glue (PVA) at a percentage of 92% Se / 8% PVA. The mixture was introduced into a syringe, which was heated with ultra-sound to obtain a paste. The desired thickness of paste was uniformly spread onto one of the backing films, then covered with a second backing film. After 10 hours of drying, the composite strip was cut to the desired length of 3546.4 mm, which incorporated 2700 mm of selenium plus additional selenium-free backing film sections for attachment to the support structure. The NEMO-3 method was used for the fabrication of 19 of the 34 selenium foils (56%). The selenium surface density for these foils ranged between 36 and 61.6 mg/cm$^2$.

– The novel method

The use of microperforated Mylar backing film can significantly affect foil radiopurity, as each fabrication step carries a risk of contamination. To mitigate this, an innovative foil–film design was developed using solid Mylar film. Initial R&D focused on identifying a radiopure mechanical structure that avoided the use of Mylar entirely. Using the BiPo-3 detector, the activities of $^{214}$Bi and $^{208}$Tl were measured for tulle—a woven nylon fabric [20]—and for raw Mylar, for comparison (Table 1).

| Material | Activity (µBq/kg) | |
|---|---|---|
| | $^{214}$Bi | $^{208}$Tl |
| PVA | <505 | <15 |
| Tulle | [275-681] | [222-407] |
| Raw Mylar | < 305 | < 65 |

Table 1: Activities of $^{214}$Bi and $^{208}$Tl measured with the BiPo-3 detector for materials considered for the novel method of source production. Limits and intervals are given at 90% C.L. [21].

When normalized to the mass of each substrate used in source fabrication - including the mass of PVA - the activities yield comparable results for both techniques, with potential for improvement in the case of Mylar, given that the values represent upper limits (table 2).

| Substrate | Activity (µBq) for 1 kg of selenium foil | |
|---|---|---|
| | $^{214}$Bi | $^{208}$Tl |
| Tulle + PVA | [60-67] | [1.7-7.8] |
| Mylar +PVA | <47 | <5.2 |

Table 2: Activities of $^{214}$Bi and $^{208}$Tl for foil substrates considered for the novel foil production, normalized to 1 kg of selenium. Limits and intervals are given at 90% C.L. [21].

The first structure studied was Nylon 6-6 fine tulle as an internal mechanical support [20]. The selected product was the lightest fabric in the Swisstulle catalogue, with a surface density of 0.7 mg/cm² and a thickness of about 0.08 mm. At SuperNEMO's request, this batch of tulle was treated with neither resins nor paint after weaving, and was only heat-set into a well defined hexagonal mesh to minimize $^{208}$Tl and $^{214}$Bi contamination. The design involved spreading a selenium–PVA paste as uniformly as possible over a thin bobbinet tulle layer. While this method offered good mechanical performance with minimal material, the absence of external protection left the foil directly exposed to the tracker gas, risking $^{82}$Se loss. In particular, PVA is sensitive to humidity, preventing the use of water in the tracker gas mixture—a possible operational modification, as in NEMO-3. Increasing the PVA content could improve foil strength but would also raise contamination levels and foil thickness. This method was therefore abandoned.

The second method used standalone selenium/PVA pads enclosed in a 12 μm raw-Mylar envelope. Initially, we followed the NEMO-3 fabrication process but replaced standard Mylar with raw Mylar; however, the lack of venting trapped water during drying. The process was then modified: selenium pads were first prepared separately, then wrapped in 12 μm raw Mylar, with intermediate welds between pads. Welding was performed by heating the Mylar with a clean soldering iron. The foil geometry—including number of pads and weld lines—was optimized to balance ease of handling (given the fragility of the selenium pads) and maximize selenium content. A configuration of eight 33 cm pads was adopted. An even number of pads was chosen because some pads developed slight curvature upon drying; alternating pads with opposing curvatures helped straighten the foil by averaging out the deformations.

### 3.3 Selenium batches

The $^{82}$Se foils were made out of several batches of $^{82}$Se that used varying purification methods. This enables a comparison between different purification methods inside the Demonstrator. Different criteria can be used to compare the methods. They can be compared for their ease of foil preparation, and for the foil radiopurity at the μBq level. This will help in future purification method selection. Six different batches were used, the characteristics of which are summarized in table 3.

| Identification | Purification method | Se Mass (kg) | $^{82}$Se (%) | Fabrication method | Number of foils |
|---|---|---|---|---|---|
| 1 | Double distillation [8] | 2.1 | 96.92-99.92 | NEMO-3 | 11 |
| 2 | Reverse chromatography [22] | 1.5 | 99.88 | novel | 7 |
| 3 | Chemical method | 0.2 | 96.9 | novel | 1 |
| 4 | Chemical method | 0.75 | 96.9 | NEMO-3 | 3 |
| 5 | Double distillation | 1.5 | 96.1 | NEMO-3 | 5 |
|   |   |   |   | novel | 2 |
| 6 | Reverse chromatography | 1 | 96.65 | novel | 5 |

Table 3: Description of the $^{82}$Se batches used for the two different methods of foil production. Batch 1 is the selenium that was recovered from NEMO-3 and reused. Batches 2 and 6 were purified recently with the novel and robust reverse chromatographic method described in [22]. They differ by the time of preparation. Batches 3, 4, and 5 were purified with more classic methods in different laboratories.

### 3.4 Foil fabrication process using the novel method

*Powder mixture preparation*
Enriched selenium was used in powder form. We first sieved the powder through a 100 μm sieve. We then ground the residue in a Retsch PM100 ball mill with a special grinding jar made out of pure iron using 20 mm diameter stainless steel balls. Mowiol 4-88 Low Ash PVA glue from Kuraray was prepared by mixing PVA flakes with ultrapure water (4.5 times weight) [23]. The cold mixture was then heated to about 70°C in a radiopure Teflon container in a water bath. We let the mixture settle for several hours (usually overnight), protected by a cover to avoid skin formation. This allowed the flakes to dissolve well, and produced the right texture.

The selenium powder and PVA mixture were weighed to a ratio of 90%/10%. The PVA proportion is slightly higher than for the NEMO-3 method with no noticeable difference in paste preparation, but with a better guarantee of the geometry in dry pad form. Selenium and PVA were mixed manually and left to settle overnight, so that the selenium agglomerates could break up in the liquid. Just before using the mixture for the foil preparation, 10% by weight of isopropanol was mixed in, creating a pasty texture and encouraging the breakage of any remaining selenium agglomerates. This had the additional advantage of creating a selenium suspension, letting the selenium remain in the mixture for longer before falling to the bottom because of its higher density. The isopropanol also facilitated quick drying. The whole process was carried out inside an ISO class 5 clean room to avoid possible radioactive contamination from the environmental atmosphere. The Se/PVA fabrication process is summarized in figure 3.

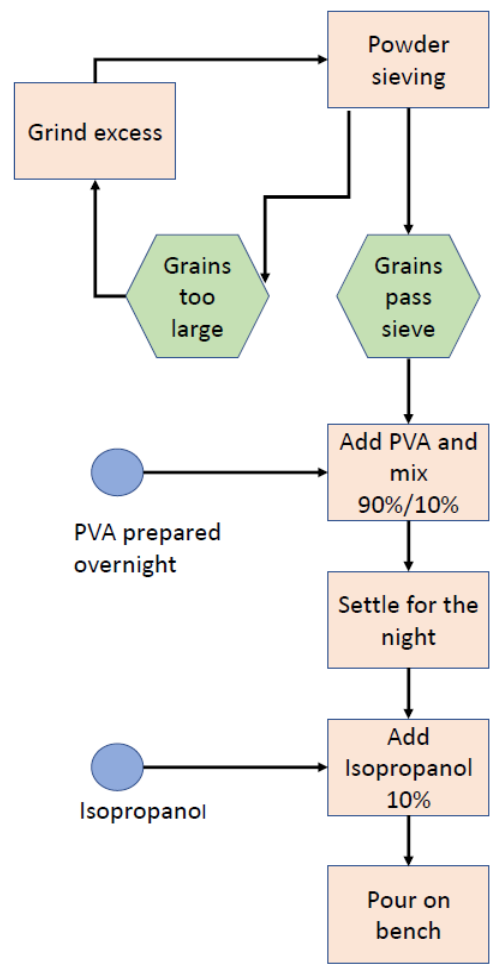

*Figure 3: Steps taken to prepare the Se/PVA mixture using the novel method.*

### Pad preparation

A film of radiopure Tedlar [24], used as an unmolding material, was stretched and placed on a bench made of Delrin. The bench was formed of removable bars to ease unmolding. In addition, a small strip of Teflon tape was glued to the bars. The homogenous Se/PVA mixture was then poured onto the Tedlar film, and spread uniformly with a calibrated scraper. Before the mixture was completely dried, the Teflon tape was removed, so that uniform longitudinal edges were formed. After drying for 24-48 hours, the removable bars were taken off, and the Se/PVA foil, still attached to the Tedlar, was released. Pads of 33x13 cm$^2$, as determined by the geometric optimization study in section 3.2, were cut out of the foil, and the Tedlar film was peeled off. Each pad was identified and weighed, the surfaces that were air-dried and Tedlar-dried were located, and the thickness was measured; all quantities were entered into a database. Particular challenges of the method included the unmolding and drying processes. With the top of the foil drying in air, and the bottom against the bench, an asymmetry in the drying pattern occurs. When this happens, the foil tends to curl up during drying, as an excess of selenium on one surface and PVA on the other generate opposing tensions in the foil. This was largely mitigated thanks to the addition of isopropanol at the Se/PVA mixing step described earlier. The resulting selenium pads had a surface density ranging from 40 to 60 mg/cm$^2$ and were about 300 µm-thick. Figure 4 shows the steps involved in the foil preparation with the novel technique.

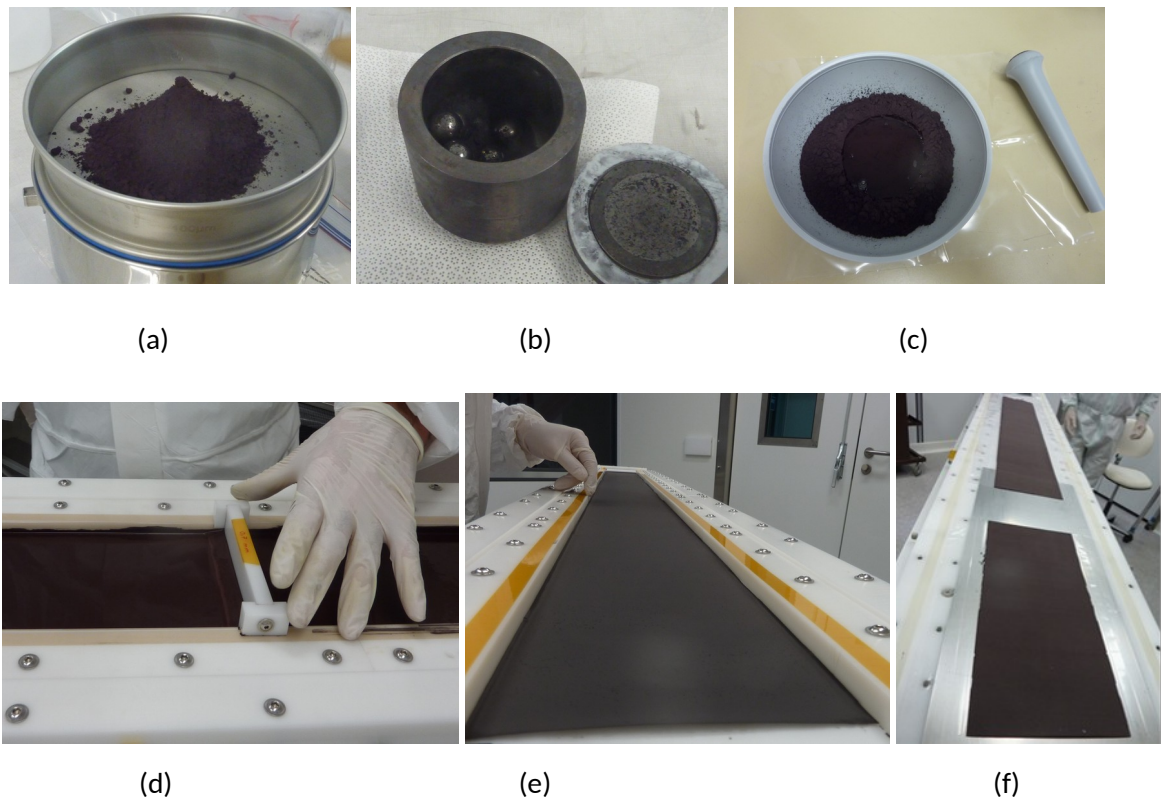

*Figure 4: Illustrations of the different foil fabrication steps. Selenium powder is first sieved (a) with the remaining powder being ground to a more homogeneous grain size (b). PVA glue and alcohol is added (c) before pouring onto a dedicated bench (d). Once dry (e), the long foil is cut into manageable 33 cm-long pads (f).*

Quanta FEG 650 Scanning Electron Microscopy (SEM) was performed on some of the selenium pads, combined with Energy Dispersive X-ray (EDX) analysis. The SEM shows a magnified view of the sample, whereas the EDX targets a specific element (Se, C) in the sample. The EDX measurements show a uniform spatial distribution of selenium and carbon atoms (in PVA) in the grains (see figure 5). All grains have a size less then 10 µm, with most smaller than 5 µm. These grain sizes are obtained for the NEMO-3 and the novel foil fabrication method.

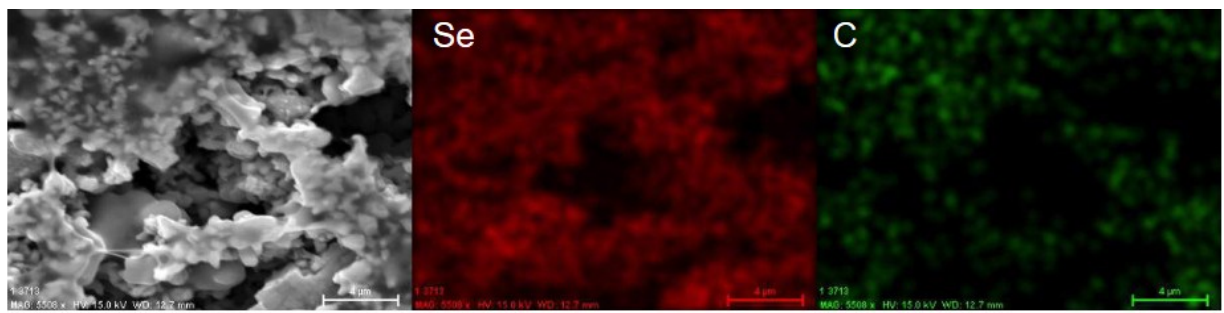

*Figure 5: Top surface (i.e. the surface in contact with air) of a selenium foil pad prepared by the novel method of foil fabrication, visualized with Quanta FEG 650 SEM, and showing selenium grains smaller than 10 μm (left); and the corresponding distributions of selenium (middle) and carbon from PVA (right), obtained with EDX.*

In addition, various $^{82}$Se batches were studied with a stereomicroscope camera Nikon SMZ800. An inappropriate grinding process for part of batch 5 (see table 3) pushed us to revert to the NEMO-3 method for its foil preparation, due to the very fine powder causing curled-up pads. It was, however, more difficult to guarantee a flat foil geometry even when using this backup process. Batch 5 powder was examined by comparing the portion that had been heavily ground with the portion subjected to less grinding. Figure 6 shows microscopy images of batch 5 in table 3: one from the portion that had only been sieved, and the other from the portion that had been heavily ground. The sieved powder shows larger grains than the heavily ground powder.

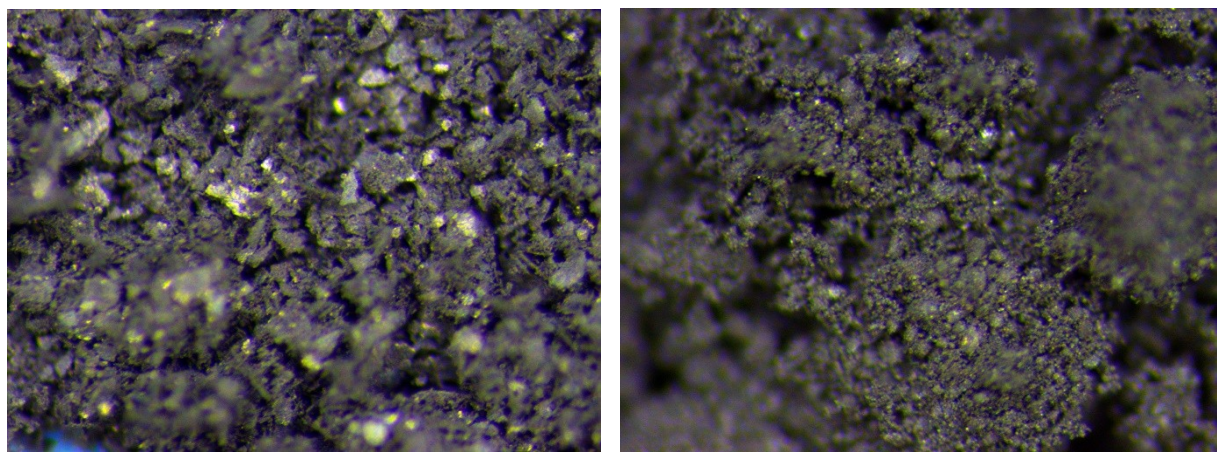

*Figure 6: Microscopy images of selenium batch 5 in table 3 from the portion that was only sieved (left), and from the portion that was heavily ground (right), magnified 6.3 times with a stereomicroscope camera Nikon SMZ800.*

The problematic batch that had been ground too heavily, and that was difficult to process, consisted of grains much smaller than those in the $^{82}$Se batches used with the novel technique, causing the PVA to form bigger clumps and the selenium to settle at the bottom. We found that by grinding the selenium to avoid agglomerates, grains and impurities larger than 100 μm, but leaving some coarser grain shapes, one can obtain a more uniform Se/PVA mixture that does not separate.

*Foil preparation*

Eight pads were chosen to form each foil. As much as possible, pads from the same selenium batch were chosen for a given foil; failing that, foils were selected that used similar purification methods. A 12 μm-thick, raw-Mylar foil was folded lengthwise in two. The first pad was inserted and the two surfaces of the Mylar foil were soldered with a dedicated clean soldering iron using a precise jig. The next pad was inserted with its top drying surface facing in the opposite direction to the previous pad, and was then soldered into the Mylar foil. This process was repeated for eight pads with alternating positioning of the top drying surface. The Mylar was then cut to obtain the correct foil length of 2700 mm of selenium, with extra Mylar at the extremities for attachment to the frame, giving a total length of 3582 mm. The selenium foil dimensions met specification tolerances, with deviations of less than 7 mm over the 2700 mm length and 0.3 mm over the 135 mm width. The foils were then folded along the soldering lines for easy packaging in plastic boxes, and stored in a clean room. Each foil took about 1.5 weeks to prepare, mainly due to the setting and drying times.

### 3.5 Copper foils

Some special foils made only from copper, known as an ultra-radiopure material, are used in SuperNEMO as control for the level of external background (i.e. external to the source foils) originating from radon daughter deposits and gamma rays interacting with the foils. These foils were made out of the metallic NEMO-3 copper foils; are 125 mm wide, 2500 mm long, with a thickness of 57.5 µm; and were prepared in the same clean room as the selenium foils (see figure 7). Some Mylar foil was glued with PVA at the extremities for installation purposes. For the nominal duration of the Demonstrator's data taking, planned at 2.5 years, it will be possible to control for the external background, using an independent measurement with the copper foils, with an accuracy of 10% [25].

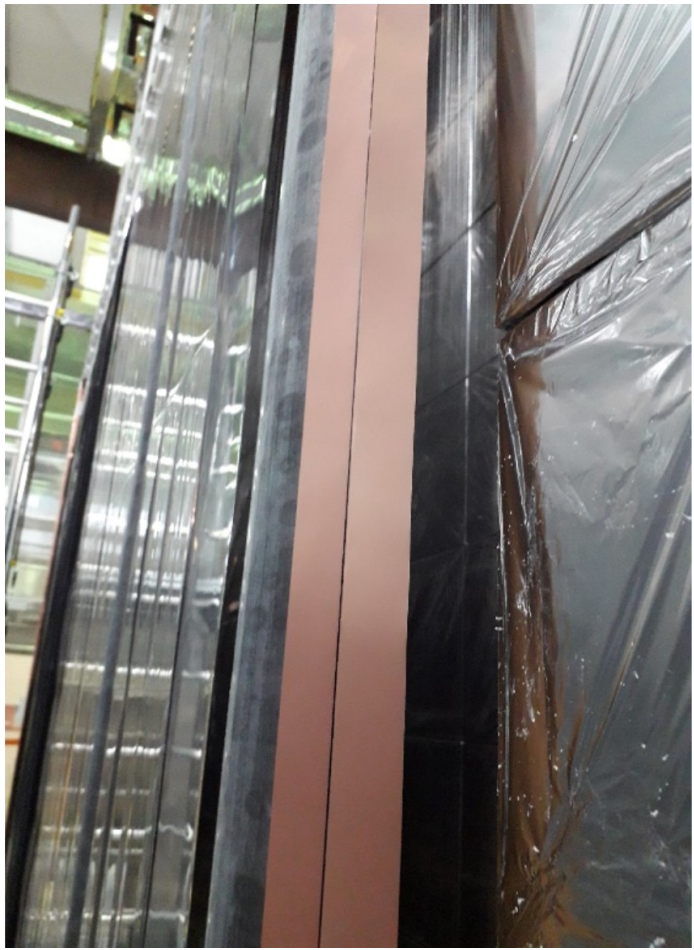

*Figure 7: One of the copper foils made out of an assembly of two ultra-pure copper strips for a total width of 125 mm.*

## 3.6 Radiopurity measurements

To meet its targets for internal backgrounds, SuperNEMO must improve on the radiopurity of the NEMO-3 source foils by an order of magnitude. The required foil radiopurities for the main background sources are $A(^{208}Tl) < 2$ µBq/kg and $A(^{214}Bi) < 10$ µBq/kg. The components of the Demonstrator foils were measured with the BiPo-3 detector at LSC [14]. It would take six months of measurements for the BiPo-3 detector to be able to achieve a sensitivity for the SuperNEMO $^{82}$Se foils of $A(^{208}Tl) < 2$ µBq/kg (90% C.L.) and $A(^{214}Bi) < 140$ µBq/kg (90% C.L.). Tables 4 and 5 give the results of activity measurements taken at BiPo-3 for the raw Mylar used in the novel method, the backing film irradiated Mylar used in the NEMO-3 foils for comparison, and the PVA. Activities are measured through two complementary methods. The first method selects alpha particles emitted by $^{212}$Po($^{210}$Po), isotopes in the decay chains of $^{208}$Tl($^{214}$Bi), with a lower energy threshold of 150 keV that reflects both bulk and surface contamination of the foils. The second method selects only alpha particles with energies below 700 keV (600 keV) in order to only measure $^{208}$Tl($^{214}$Bi) bulk contamination.

| Sample | Mass (g) | Time (days) | S (m$^2$) | $^{208}$Tl Activity (µBq/kg) 90% C.L. |
|---|---|---|---|---|
| **Raw Mylar** | 108.1 | 76.5 | 1.62 | |
| E($\alpha$)>150 keV | | | | <65 |
| 150<E($\alpha$)<700 keV | | | | <49 |
| **Backing film Mylar (irr.)** | 200 | 44.4 | 3.06 | |
| E($\alpha$)>150 keV | | | | $98^{+62}_{-47}$ |
| 150<E($\alpha$)<700 keV | | | | $90^{+63}_{-42}$ |
| **PVA** | 210 | 137.2 | 1.8 | |
| E($\alpha$)>150 keV | | | | <15 |
| 150<E($\alpha$)<700 keV | | | | <12 |

Table 4: Results of $^{208}$Tl activity measurements with BiPo-3 of raw Mylar, backing film and PVA. Time is the duration of the measurement, S is the active detection area. Different selections based on the energy of emitted alpha particles (from $^{212}$Po in the $^{208}$Tl decay chain) have been applied in the analysis to identify bulk contamination from surface contamination [14].

| Sample | Mass (g) | Time (days) | S (m$^2$) | $^{214}$Bi Activity (µBq/kg) 90% C.L. |
|---|---|---|---|---|
| **Raw Mylar** | 108 | 31.1 | 1.62 | |
| E($\alpha$)> 300 keV | | | | <305 |
| 300<E($\alpha$)<600 keV | | | | <195 |
| **Backing film Mylar (irr.)** | 190 | 31.3 | 2.88 | |
| E($\alpha$)>300 keV | | | | - |
| 300<E($\alpha$)<600 keV | | | | <688 |
| **PVA** | 230 | 124.6 | 1.8 | |
| E($\alpha$)>300 keV | | | | - |
| 300<E($\alpha$)<600 keV | | | | <505 |

Table 5: Results of $^{214}$Bi activity measurements with BiPo-3 of raw Mylar, backing film and PVA. Time is the duration of the measurement, S is the active detection area. Different selections based on the energy of emitted alpha particles (from $^{210}$Po in the $^{214}$Bi decay chain) have been applied in the analysis to identify bulk from surface contamination [14].

As expected, the irradiated Mylar appears to be more contaminated (at least for $^{208}$Tl, where a positive value is observed) since it was exposed to radiation followed by chemical etching steps. These additional steps could add extra impurities.

During production and before installation, the foils destined for the Demonstrator were measured in BiPo-3. Due to lack of time, not all selenium batches were measured with BiPo-3. Table 6 shows the global results of $^{208}$Tl and $^{214}$Bi measurements of sample foils to validate the production methods from the batches with the highest mass inside the Demonstrator and with various purification methods.

| Se Batch | Se Mass (kg) | Se powder purification | Foil production method | A($^{208}$Tl) (µBq/kg) | A($^{214}$Bi) (µBq/kg) |
|---|---|---|---|---|---|
| 1 | 1.95 | Double distillation | NEMO-3 | 24[13-38] | <290 |
| 2 | 1.5 | Reverse chromatography | Novel | 22[8-54] | <595 |
| 5 | 1.4 | Double distillation | Novel | 131[63-243] | <525 |
| 6 | 0.97 | Reverse chromatography | Novel | <106 | <1374 |

Table 6: Results of $^{208}$Tl and $^{214}$Bi measurements with 90% C.L. intervals and limits at 90% C.L. for selenium ββ foil samples from the batches with the highest mass inside the Demonstrator.

Due to the low mass of the measured samples, only upper limits were able to be set for some of the activities. None of the foils were measured for the 6 months required for the maximum sensitivity of BiPo-3, but only for approximately half that time. For the measurements that detect $^{208}$Tl activity as opposed to limits, batches 1 and 2 give the best radiopurity results, the first being the double-distilled $^{82}$Se used in NEMO-3, and the second being that which was purified with the novel reverse-chromatographic method. These two batches also have the highest enrichment (see table 3). Longer measurements with the SuperNEMO Demonstrator on a larger number of foils will allow us to draw conclusions on the best method from the point of view of radiopurity, and to check for possible isolated hot spots. Nevertheless, if these values are confirmed, $^{208}$Tl radiopurity will be at least five times better than that of the composite molybdenum foils in NEMO-3, and fifteen times better than the selenium radiopurity in NEMO-3 [26].

Since BiPo-3 is also sensitive to surface contamination, a comparison was made between the radiopurities of the top and bottom surfaces, revealing no differences despite the fact that during foil drying, one surface was in contact with air and the other was in contact with the unmolding Tedlar. This shows that the preparation method was undertaken with sufficiently radiopure equipment, in a well-controlled environment.

In addition to BiPo-3 measurements, we also made ICP-MS measurements for comparison. The apparatus used was an Agilent 7900 ICP-MS in an ISO class 6 room. This method enables the detection of $^{238}$U and $^{232}$Th contamination, the parent isotopes of $^{214}$Bi and $^{208}$Tl, respectively. Two selenium samples were studied. They were purified with the same method as batch 2: one was natural selenium, the second one was enriched $^{82}$Se. The results are summarized in table 7.

| Sample | $^{232}$Th (µBq/kg) | $^{238}$U (µBq/kg) |
|---|---|---|
| Natural-Se-like batch 2 | 130±20 | 274±37 |
| Enriched-Se batch 2 | 61±24 | 112±75 |

Table 7: ICP-MS results for two selenium powder samples.

Using these values and assuming equilibrium, the value for $^{208}$Tl in the enriched sample is 22 ± 9 µBq/kg. This value can be compared to the $^{208}$Tl activity obtained in BiPo-3 for a similar sample in table 6 that is 22 [8-54] µBq/kg. Two different radiopurity measurement methods showed the same results for several samples purified by the same method. We are thus confident in the radiocontaminant analyses we carried out using the BiPo detector. Locally, the radiopurity may be slightly higher, pointing to contaminants concentrated in hot spots. The effect of these can be removed by post-analysis.

## 3.7 Foil integration

*Transport*

The source foils were prepared either with the novel method or the NEMO-3 method, and packed in a clean room before delivery to LSM. The foils were wrapped in additional mylar, sealed in tight plastic bags, and stored in clean boxes. These boxes were then wrapped with food-grade film for transport. Finally, they were delivered to LSM inside the integration clean tent housing the demonstrator under construction.

*Installation*

All foils have Mylar ends with holes for attaching to a specific support on the frame that can be hooked into place and correctly tensioned, allowing for a close alignment of the selenium surface of the foil directly parallel to the calorimeter walls. The foils were installed between the two tracker halves while the Demonstrator was open. The space between foils was minimized to ensure maximum surface coverage. Between every sixth foil, a space of 18 mm accommodates the $^{207}$Bi sources of the calibration system. Figure 8 shows the foils after installation, where one can distinguish the NEMO-3 foils, produced in a single continuous piece, from the foils made using the novel method, which feature eight pads. The different purification methods produce different coloured powders: a dark grey colour for batches 1, 3, 4 and 5 in table 3, and a reddish colour for batches 2 and 6 in table 3. In addition, for the novel method, since the pads are alternated in their orientation (between the top and bottom surfaces of the pads during production), the surface visible in the picture renders different shades of red colour even if the selenium preparation is the same. In particular, the bottom surface, which was in contact with the unmolding Tedlar during production, is smoother than the top surface. The source foils were installed on the source frame inside the clean tent using dedicated manipulation tools to ensure cleanliness and integrity. After installation, the demonstrator was closed by joining its two halves, thereby protecting the foils from exposure and damage.

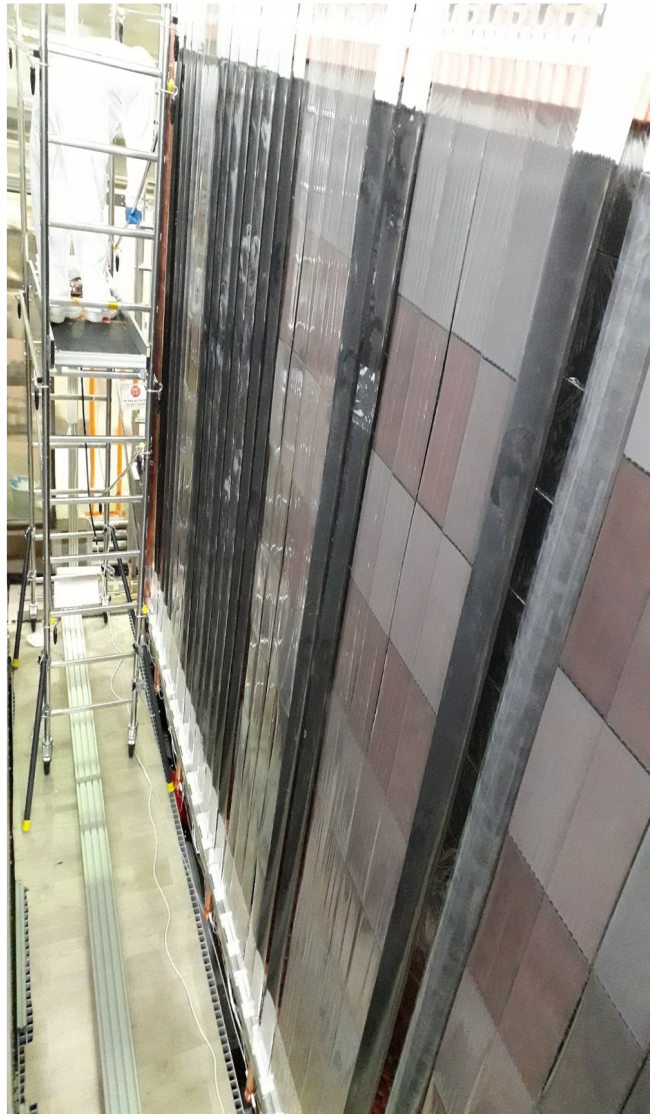

*Figure 8: SuperNEMO ββ foils inside the Demonstrator before complete closing. One can distinguish the foils made using the NEMO-3 method, in one piece (with a dark grey colour from batches 1, 3, 4 and 5 in table 3,) from the foils made using the novel method, which feature eight pads (with a reddish colour from batches 2 and 6 in table 3) .*

### Foil geometry

The exact positions and geometry of foils installed in the Demonstrator were measured using laser tracking with a Leica T-Scan 5 with a fraction-of-a-millimeter resolution. Figure 9 shows the global survey of all the foils, where each blue point represents a measurement point. One can see both types of foils, and can distinguish the foils prepared with the NEMO-3 method from those manufactured with the novel technique by their planar geometry. The $^{207}$Bi calibration sources, in red, were deployed between the foils during the laser scan. The two, shorter copper foils at each extremity are also visible. The selenium sections of the foils have a nominal length of 2700 mm, and a width of 135 mm, the total width covered by the foils being 5000 mm. The top part of the figure shows a close-up of the transverse profile (top view) to highlight the shape features.

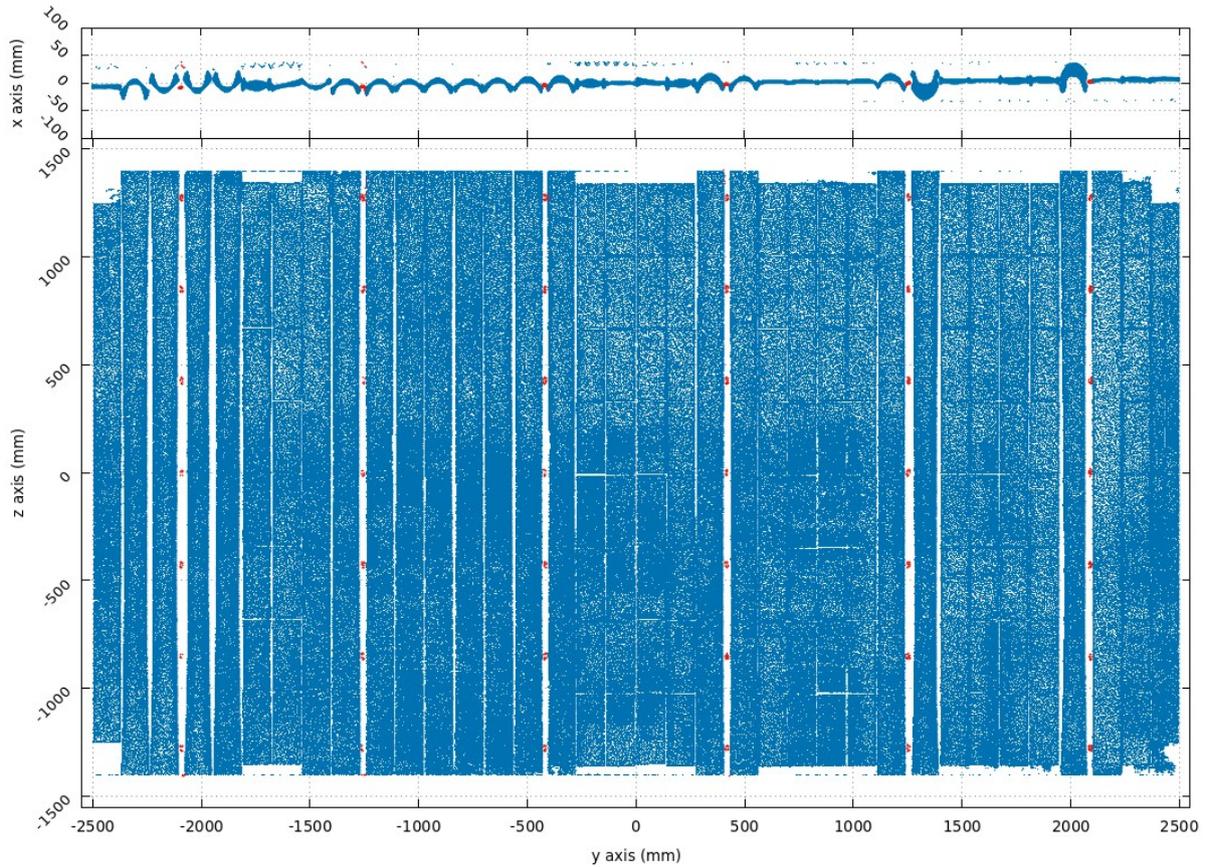

*Figure 9: Laser scan of SuperNEMO foils inside the Demonstrator, with dimensions, before final closure of the detector. On top, the corresponding foils are seen from above. At the bottom, we see the foils from the front. The red spots indicate the $^{207}$Bi calibration sources, which were deployed during the laser scan.*

Foils produced by the NEMO-3 method exhibit some curvature, with top and bottom extremities being less curved than the centre. The foils made with the novel method are globally flatter. This is due to the mechanical constraint of the regularly-spaced transverse soldering that stiffens the foil, in addition to the fact that the pads were soldered with alternate orientations with respect to the surface that was on top during the drying process on the bench. Figure 10 shows the front view scan of a foil made with the NEMO-3 method, together with a foil from the novel method, in more detail. The slight difference in length comes from the fabrication process with the novel method cutting the pads to the right size, whereas the NEMO-3 method pours directly onto the backing film. The illusion of a narrow waist of the foil is a consequence of the curvature as seen in projection.

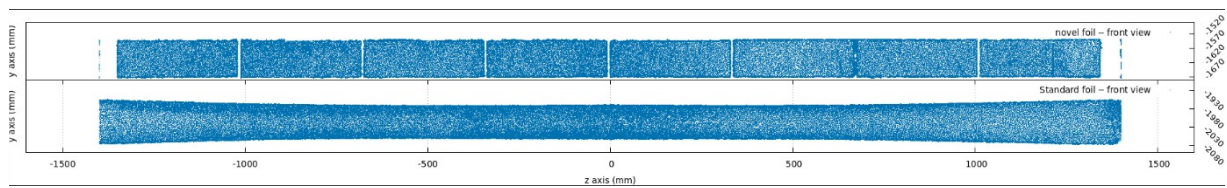

*Figure 10: Laser scan close-up of a foil made using the NEMO-3 method (below) and a foil made using the novel method with eight pads (above), inside the Demonstrator, before final closure of the detector.*

It is currently unknown if the foils will preserve this geometry over time. In NEMO-3, some foil deformation, not present at the beginning of the experiment, was observed upon opening and dismantling the detector, 10 years after installation. An example foil prepared by the novel method has been kept in our clean room for several years without any noticeable changes. After some initial shrinkage of the foils due to additional drying of stored pads awaiting Mylar wrapping, the foils prepared by the new method did not change shape. In addition, we performed an aging test by installing a short pad sealed in Mylar foil into a test chamber flushed by the SuperNEMO gas mixture of helium, ethanol, and argon (95% He+4% $C_2H_5OH$+1% Ar) for several months, without any significant change of shape. This exercise was performed only for the novel foil fabrication method, the NEMO-3 foils having already been tested in real conditions for several years.

Realistic foil geometries have been implemented into Falaise, the SuperNEMO software toolkit [27]. Figure 11 shows the Demonstrator response from a simulated electron generated in the bulk of a curved foil crossing the tracker chamber and hitting a calorimeter optical module. A preliminary study suggests a loss of efficiency in the two-electron channel of the order of 2% due to the curvature. More detailed study is planned to handle the effect in future data analysis.

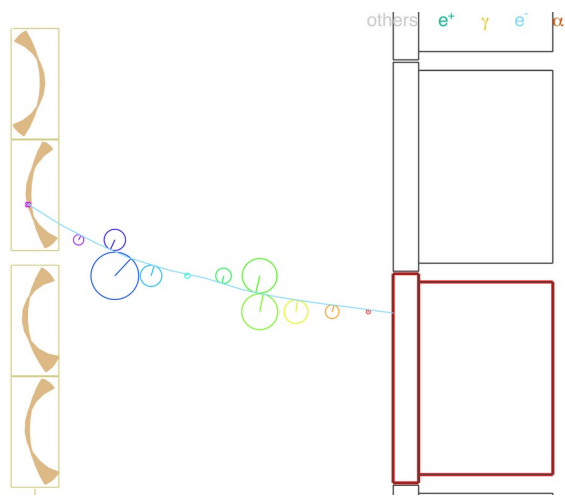

Figure 11: GEANT4-based simulation [28] of a 1 MeV electron in the SuperNEMO Demonstrator (zoomed view) with a vertex (pink cross) on the foil made using the NEMO-3 method (brown curves) and a hit in a calorimeter optical module (red box). The coloured circles represent the tangent radius of the electron's path, deduced from the signal induced in the Geiger cells. The simulated electron trajectory, curved by the 25 G magnetic field, is shown in cyan.

# 4. Conclusion

The SuperNEMO Demonstrator incorporates a ββ decay source consisting of independent selenium source foils in the middle of the detector. Different purification processes, and two foil fabrication methods, were used to produce 34 foils, 2.7 m long, 13 cm wide, and about 300 µm-thick, using 6.11 kg of enriched selenium. We have shown that the methods provide radiopure selenium foils. In particular, promising results have been obtained for the novel foil-fabrication method consisting of wrapping standalone selenium pads in raw Mylar, and for the selenium purified by the new reverse-chromatography method. The new foils have been measured to be a factor of 15 times more radiopure than the NEMO-3 selenium foils. In addition, the novel method, using alternating pad orientations and strengthening with soldering lines, suggests an improved potential to maintain a flat profile over time. This will be further studied by long-term analysis.

The source foils have been installed and the Demonstrator has been fully assembled. Data taking started in early 2025 with the aim of demonstrating the capability of a large scale tracker-calorimeter experiment to search for 0νββ with $^{82}$Se.

## 5. Acknowledgements

The authors wish to thank the Modane Underground Laboratory and its staff for support through underground space, logistical and technical services. LSM operations are supported by the CNRS, with underground access facilitated by the Société Française du Tunnel Routier du Fréjus.

The authors wish to thank Kuraray for providing different samples of PVA and sharing their experience before choosing their Mowiol 4-88 Low Ash.

The authors also wish to thank Daniel Redman UT Chemistry who contributed to SEM images and EDX analysis.

The authors wish to thank Jean-Marc Dubois from LAPP for his precious help and ingenious ideas in elaborating the novel foil preparation technique.

Finally, we acknowledge support by the grants agencies of the CNRS/IN2P3 in France, the MEYS of the Czech Republic (LRI LSM-CZ), NSF in the USA, Slovak Research and Development Agency (projects APVV-15-0576 and APVV-21-0377) and the Science and Technology Facilities Council (STFC), part of United Kingdom Research and Innovation (UKRI).